# Isotropic fabrication of centimeter-scale, low propagation-loss periodically poled lithium niobate nanophotonic waveguides for efficient second harmonic generation

Guanghui Zhao,[1,4,†] Yixuan Yang,[1,5,†] Renhong Gao,[2,3] Jintian Lin,[1,5,§] and Ya Cheng[2,3,6,7,8,9,*]

[1]*State Key Laboratory of Ultra-intense Laser Science and Technology, Shanghai Institute of Optics and Fine Mechanics, Chinese Academy of Sciences, Shanghai 201800, China*
[2]*State Key Laboratory of Precision Spectroscopy, East China Normal University, Shanghai 200062, China*
[3]*The Extreme Optoelectromechanics Laboratory (XXL), School of Physics, East China Normal University, Shanghai 200241, China*
[4]*School of Physical Science and Technology, ShanghaiTech University, Shanghai 200031, China*
[5]*Center of Materials Science and Optoelectronics Engineering, University of Chinese Academy of Sciences, Beijing 100049, China*
[6]*Shanghai Research Center for Quantum Sciences, Shanghai 201315, China*
[7]*Hefei National Laboratory, Hefei 230088, China*
[8]*Collaborative Innovation Center of Extreme Optics, Shanxi University, Taiyuan 030006, China*
[9]*Collaborative Innovation Center of Light Manipulations and Applications, Shandong Normal University, Jinan 250358, China*
[†]*These authors contributed equally to this work*
[§]*E-mail: jintianlin@siom.ac.cn*
**E-mail: ya.cheng@siom.ac.cn*

**Abstract:** Periodically poled lithium niobate (PPLN) nanophotonic waveguides that simultaneously feature low propagation-loss and uniform periodic poling are essential for a wide range of applications ranging from classical nonlinear frequency-conversion to scalable integrated quantum technology. However, fabrication imperfections have frequently limited the propagation loss of fully domain-inverted PPLN nanophotonic waveguides to a few dB/cm, primarily due to anisotropic etching issue, thereby restricting the absolute conversion efficiency and scale of photonic integration. Here, we present a fabrication approach that overcomes this challenge, yielding a 1.2-cm-long PPLN nanophotonic waveguide with low propagation loss via femtosecond-laser photolithography-assisted chemo-mechanical etching (PLACE). By carrying out domain inversion on a planar thin-film prior to waveguide definition, electric-field distortion is minimized during poling, while isotropic etching of the waveguide is achieved by PLACE with an average surface roughness of only 0.34 nm, resulting in uniform poling of duty cycle of 50% and a record-low propagation loss of 0.042 dB/cm in the telecom band. Under continuous-wave pumping at 1525 nm, the device demonstrates a high normalized quasi-phase-matched SHG conversion efficiency of 2021% $W^{-1}$, and an absolute conversion efficiency of 64% at a pump power of 86 mW which represents the state of the art for single-period PPLN nanophotonic waveguides.

## 1. Introduction

Thin-film lithium niobate (TFLN) on insulator has emerged as a versatile platform for photonic integration with unprecedented performance, owing to its exceptional combination of a wide transparency window, moderate refractive-index contrast, strong electro-optic response, and large second-order nonlinearity.[1–5] Rapid advances in nanofabrication have enabled TFLN nanophotonic waveguides with compact cross-sections and propagation loss as low as <1 dB/m in the telecom band,[6-8] thereby providing strong optical confinement and high integration density. The potential of this platform is evidenced by a variety of integrated photonic devices, including ultra-high speed electro-optic modulators,[9,10] bi-chromatic soliton microcombs,[11-13] broad-bandwidth supercontinuum generation,[14] narrow-linewidth microlasers,[15-17] and highly efficient nonlinear frequency convertors.[18-23] In particular, periodically poled lithium niobate (PPLN) nanophotonic waveguides on the TFLN platform dramatically enhance nonlinear frequency conversion by simultaneously leveraging the largest second-order nonlinear coefficient (i.e., $d_{33}$~27 pm/V),[14,17,19,20,23-26] tight optical confinement, and low propagation loss. Accordingly, low-loss PPLN nanophotonic waveguides have been extensively employed for boosting classical nonlinear processes such as second harmonic generation (SHG),[17,23-26] sum-frequency generation,[19] and optical parametric oscillation,[20] substantially extending the spectral coverage of coherent light sources. Beyond classical frequency conversion, PPLN nanophotonic waveguides have also become key building blocks for quantum photonics, enabling entangled photon-pair generation,[27-30] squeezed light generation,[31] and even single-photon-level nonlinearities for quantum photonic circuits.[32–34] All these nonlinear applications demand both low propagation loss and uniform poling—i.e., complete domain inversion across the waveguide cross-section with an optimum 50% duty cycle—to maximize the effective interaction length and thereby enhance the nonlinear interaction enabled by quasi-phase matching.[35]

The fabrication of PPLN nanophotonic waveguides involves two essential steps: waveguide etching and ferroelectric domain inversion, which can be performed in either order. However, these two processes are often interdependent, leading to a tradeoff between propagation loss and poling uniformity. If the ridge waveguide is etched before high-voltage poling, anisotropic etching of the waveguide is significantly suppressed, so a low propagation loss of 0.25 dB/cm can be achieved in the telecom band.[36] However, the local electric fields become susceptible to perturbations from sidewalls and surface topography, which can degrade domain uniformity.[37,38] Conversely, if the ridge waveguide is fabricated after poling, the conventional dry-etching and subsequent chemical cleaning processes induce anisotropic, domain-dependent surface relief, period distortion, and additional scattering loss, because oppositely poled domains exhibit different etching responses. This typically results in a propagation loss of several dB/cm in the telecom band (e.g., 3 dB/cm in Ref. [23]). Such imperfections are especially detrimental for centimeter-long PPLN waveguides aimed at high absolute conversion efficiency by using quasi-phase matching (QPM) scheme, where weak scattering and small poled-period deviations accumulate over the entire interaction length, progressively increasing scattering loss and causing power to flow back from the generated nonlinear signals to the pump, thereby depleting the nonlinear output and limiting the scalability of photonic integration.

In this work, we overcome these challenges and demonstrate a 1.2-cm-long PPLN nanophotonic waveguide featuring uniform poling and a record-low propagation loss of 0.042 dB/cm in the telecom band, by leveraging femtosecond laser photolithography assisted chemo-mechanical etching (PLACE). The key advantage of this fabrication strategy lies in both uniform domain inversion of the planar TFLN by high-voltage poling and patterning the pre-poled TFLN into PPLN waveguides with ultra-smooth surfaces by chemo-mechanical polishing (CMP) rather than by a strongly anisotropic dry-etching process, yielding ultra-efficient quasi-phase-matched SHG with an absolute conversion efficiency as high as 64% at an on-chip pump power of 86 mW. This work unlocks the full potential of PPLN nanophotonic waveguides for a wide range of nonlinear photonic applications.

## 2. Fabrication and Characterization of Centimeter-Scale Low-Loss PPLN Nanophotonic Waveguides

### 2.1 Device Fabrication and Process Strategy

The devices were fabricated on a commercial X-cut TFLN wafer composed of a 500-nm-thick TFLN device layer, a 4.7-μm-thick buried $SiO_2$ layer, and a 500-μm-thick lithium niobate handle. The fabrication flow is schematically shown in Fig. 1. First, a 200-nm-thick chromium (Cr) film was deposited on the TFLN wafer by magnetron sputtering. Periodic microelectrodes along the Y axis of the lithium niobate crystal were then patterned in the Cr film by spatially selective femtosecond-laser ablation with a resolution of approximately 100 nm.[39] High-voltage poling was subsequently applied to periodically pole the X-cut TFLN using a pre-pulse followed by a main pulse. The pre-pulse voltage consisted of four triangular pulses with a voltage amplitude of 320 V, each having a rise time and a fall time of 0.5 ms. These pre-pulses generate sufficient domain nucleation sites near the positive electrode while suppressing excessive lateral broadening of the domain structures. After the pre-pulses, four main pulses with a peak voltage of 320 V and a peak duration of 6 ms were applied to drive the reversed domains to grow through the film thickness, thereby determining the final domain width and duty cycle. This separation of domain nucleation and domain growth facilitates high-fidelity periodic poling with a duty cycle close to 50%.

After poling, the residual Cr electrodes were removed and the sample was cleaned. Ridge waveguides were then isotropically defined in the periodically poled TFLN regions via the PLACE technique which consisted of five steps.[8] First, a fresh Cr film was deposited as a hard mask for waveguide fabrication. Second, stripe patterns were written in the Cr film by spatially selective femtosecond-laser ablation. Third, the unprotected LN (without Cr mask coverage) was removed by CMP, producing a smooth ridge waveguide without inducing anisotropic etching. Fourth, the residual Cr mask was stripped. Finally, a 1.5-μm-thick $SiO_2$ upper cladding was deposited by plasma-enhanced chemical vapor deposition, and the chip facets were polished by CMP for end-fire coupling. This fabrication sequence avoids applying the poling field to a pre-etched topography and employs CMP to suppress surface scattering loss.

**2.2 Structural Characterization and Propagation Loss Measurement**

The fabricated PPLN ridge waveguide features an effective poled waveguide length of 1.2 cm, a ridge top-width of approximately 1 μm, and an etch depth of around 250 nm. The ridge waveguide is covered with a 1.5-μm-thick $SiO_2$ cladding that symmetrizes the optical environment, improves the stability of the phase-matching condition, and protects the waveguide from contamination. The optical micrograph in Fig. 2(a) confirms a straight and continuous ridge waveguide over the imaged region, and the cross-sectional scanning electron microscope (SEM) image in Fig. 2(b) verifies that the ridge waveguide is embedded in the $SiO_2$/TFLN/$SiO_2$ stack. Atomic-force microscopy (AFM) performed on the waveguide surface before depositing the $SiO_2$ cladding yields a root-mean-square (RMS) roughness of only 0.34 nm (Fig. 2(c)), demonstrating that the CMP process produces a sub-nanometer-smooth surface suitable for centimeter-scale low-loss propagation. To quantify the propagation loss, a racetrack microring resonator with a physical cavity length of 1828 μm (= 2π×100 + 600×2) fabricated on the same wafer with the identical poling and CMP process was characterized, where one straight section of the racetrack was periodically poled with a length of 600 μm. Fitting of the resonance with a Lorentz curve gives an intrinsic optical quality factor of $7.98\times10^6$, corresponding to an estimated propagation loss of 0.042 dB/cm (Fig. 2(d)). This propagation loss represents an approximately two-order-of-magnitude improvement over the results achieved by conventional pole-before-etch-process in TFLN.[23,26] This result confirms that the post-poling CMP process preserves low-loss guiding even in periodically poled regions.

### 2.3 Quasi-Phase-Matching Design and Domain Characterization

For first-order QPM in type-0 SHG, the phase-matching (i.e., momentum conservation) condition is[35]

$$\Delta k = k_{2\omega} - 2k_{\omega} = 2\pi/\Lambda, \tag{1}$$

where $\Lambda$ is the poling period. Equivalently, for a pump wavelength $\lambda_{\omega}$,

$$\Lambda = \lambda_{\omega}/[2(n_{2\omega} - n_{\omega})]. \tag{2}$$

Finite-element mode simulations were employed to calculate the modal effective indices and determine the required QPM period. As shown in Fig. 3(a), the fundamental wave near 1525 nm and the second-harmonic wave near 762.5 nm are both $TE_{00}$-like modes, with effective indices of 1.904 and 2.091, respectively. According to Eq. (2), first-order QPM period was designed to be 4.13 μm for $TE_{00}$-to-$TE_{00}$ SHG near 1525 nm. Using of the same spatial-mode order for both fields is advantageous for the mode overlap and helps suppress conversion into higher-order modes. The influence of the $SiO_2$ upper cladding was then evaluated by calculating the QPM period as a function of cladding thickness. The period varies rapidly for thin cladding but saturates when the $SiO_2$ thickness exceeds approximately 0.8 μm (Fig. 3(b)), indicating that the upper boundary is sufficiently far from the optical mode. Consequently, a 1.5-μm-thick $SiO_2$ cladding was chosen to minimize the sensitivity to deposition non-uniformity. Figure 3(c) further displays the calculated QPM period versus pump wavelength under this cladding condition.

The ferroelectric domain quality was characterized using second-harmonic microscopy. An image taken over a 100-μm-long region (Fig. 3(d)) reveals clear periodic contrast, continuous domain inversion along the propagation direction, with a duty cycle close to 50:50 near the waveguide path. These observations confirm that the pre-pulse/main-pulse poling waveform and the planar-film poling geometry produce sufficiently uniform domains for the 1.2-cm-long QPM interaction.

**2.4 Second-Harmonic Generation Characterization**

The experimental setup for SHG characterization is illustrated in Fig. 4. A tunable continuous-wave laser in the telecom band was used as the pump source, and its output was amplified by an erbium-doped fiber amplifier (EDFA) when high-power measurements were required. A fiber polarization controller was employed to adjust the input state to transverse-electric (TE) polarization, so that the pump field could access the largest nonlinear tensor component $d_{33}$ through type-0 QPM. The pump light was coupled into the PPLN nanophotonic waveguide by a lensed fiber, and the transmitted pump and generated second-harmonic (SH) signals were collected by another lensed fiber at the output facet. The output was sent to an optical spectrum analyzer (OSA) to record the fundamental and SH spectra. A top-view microscope imaging system was simultaneously used to monitor fiber-waveguide alignment and visible SH emission from the waveguide, which provides a direct indication of the coupling condition and frequency-conversion strength. The pump and SH powers reported below are on-chip values after correction for the corresponding coupling losses. Low-power pump-wavelength scans were performed to extract the device normalized conversion efficiency while minimizing thermal drift and photorefractive effects, so that the measured spectral response more faithfully reflects the intrinsic QPM conversion behavior of the device. The pump power was then increased stepwise for the absolute-efficiency measurement.

At low pump power, the normalized SHG efficiency was extracted as

$$\eta_{norm} = P_{2\omega}/P_{\omega}^{2}, \tag{3}$$

where $P_\omega$ and $P_{2\omega}$ are the on-chip pump and SH powers, respectively. This definition represents the device conversion efficiency without normalization to the interaction length squared. In the low-power measurement, the pump wavelength was scanned around the designed QPM wavelength while the on-chip pump power was kept at approximately 0.5 mW. The normalized conversion-efficiency spectrum reaches a maximum of 2021% $W^{-1}$ near 1525 nm (Fig. 5(a)). The measured response follows a $sinc^2$-shaped envelope with a full width at half maximum of approximately 1 nm. Such a narrow and nearly ideal spectral response is consistent with coherent SHG over the full 1.2-cm effective poled length and indicates that the QPM period, ferroelectric duty cycle, and waveguide geometry remain sufficiently uniform along the centimeter-scale device.

At higher powers, the absolute conversion efficiency was defined as

$$\eta_{abs} = P_{2\omega}/P_{\omega}. \quad (4)$$

The SH output power increased monotonically with increasing on-chip pump power, as shown in Fig. 5(b). When the on-chip pump power was increased to 50.93 mW, the measured SH power reached 23.10 mW, corresponding to an absolute efficiency of 45.4% (Fig. 5(c)). At an on-chip pump power of 86.09 mW, the SH output further increased to 55.60 mW, yielding a peak absolute efficiency of 64.6% (Fig. 5(d)). To the best of our knowledge, these values represent the highest device normalized conversion efficiency and absolute conversion efficiency reported for a single-period PPLN nanophotonic straight waveguide.

Because no resonant enhancement was used in these measurements, the high efficiency originates from the combined effects of long interaction length, low propagation loss, uniform QPM, and strong $TE_{00}$-$TE_{00}$ modal overlap. The simultaneous observation of a $sinc^2$-like wavelength response and high absolute conversion efficiency confirms that the device operates as a uniform single-period PPLN frequency converter rather than relying on localized enhancement or accidental phase matching.

## 3. Discussion and Conclusion

### 3.1 Discussion

It is worth noting that the present process is distinct from the previously reported pole-before-etch routes.[26,36,37] The key advantage lies in patterning the pre-poled TFLN by CMP rather than by a strongly anisotropic dry-etching step. During high-voltage poling, the unetched TFLN provides a nearly planar surface and a more uniform local electric field; in contrast, poling performed on an already etched ridge structure would lead to electric-field crowding near sidewalls and steps, causing nonuniform domain nucleation, excessive domain broadening, and even duty-cycle errors. Once the domain pattern is established, CMP defines the ridge waveguide through a nearly isotropic material-removal process, thereby eliminating the domain-dependent etching contrast between the inverted and non-inverted regions. This combination effectively decouples domain formation from low-loss waveguide definition, enabling efficient quasi-phase matching (QPM) together with sub-nanometer surface roughness in a centimeter-scale device.

### 3.2 Conclusion

We have demonstrated a centimeter-scale PPLN ridge waveguide fabricated by femtosecond-laser photolithography-assisted chemo-mechanical etching. By carrying out domain inversion on a planar film prior to waveguide definition, the process minimizes electric-field distortion during poling, while CMP provides a sub-nanometer-roughness waveguide surface. The device has a 1.2-cm effective poled length and a designed QPM period of 4.13 μm for $TE_{00}$-to-$TE_{00}$ SHG near 1525 nm. It exhibits an RMS roughness of 0.34 nm, an intrinsic Q factor of $7.98\times10^6$ measured from a same-process microresonator, and a propagation loss of 0.042 dB/cm. The device achieves a normalized SHG efficiency of 2021% $W^{-1}$ and an absolute conversion efficiency of 64% at an on-chip pump power of 86.09 mW. This fabrication strategy provides a scalable route toward low-loss, high-efficiency TFLN nonlinear photonic circuits.

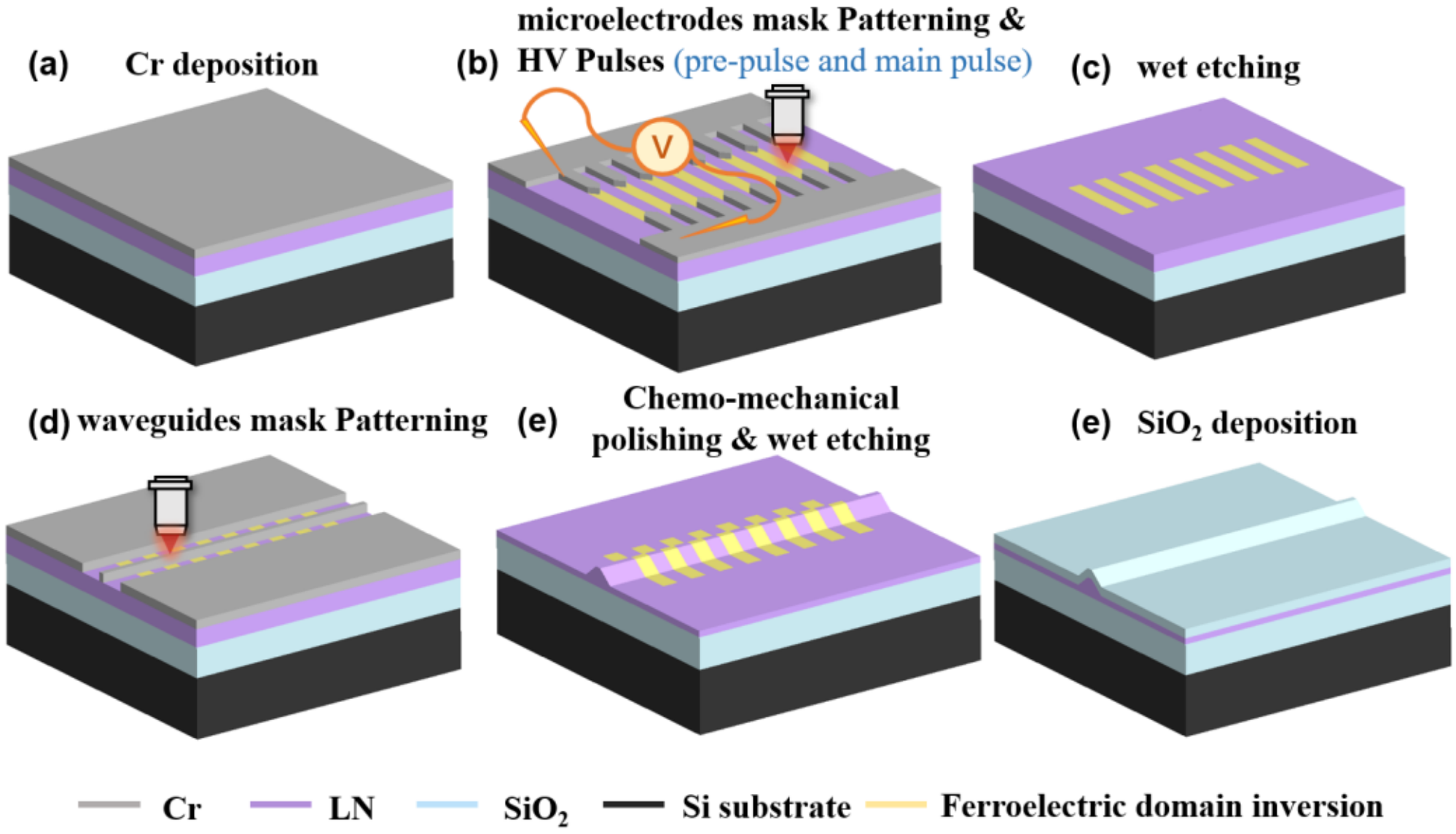

**Figure 1.** Fabrication process of the PPLN nanophotonic ridge waveguide. A Cr film is deposited and patterned into periodic poling electrodes. After high-voltage poling and electrode removal, a second Cr mask is written for waveguide definition. Chemo-mechanical polishing (CMP) forms the ridge waveguide by etching the unprotected lithium niobate, followed by Cr removal and $SiO_2$ cladding deposition.

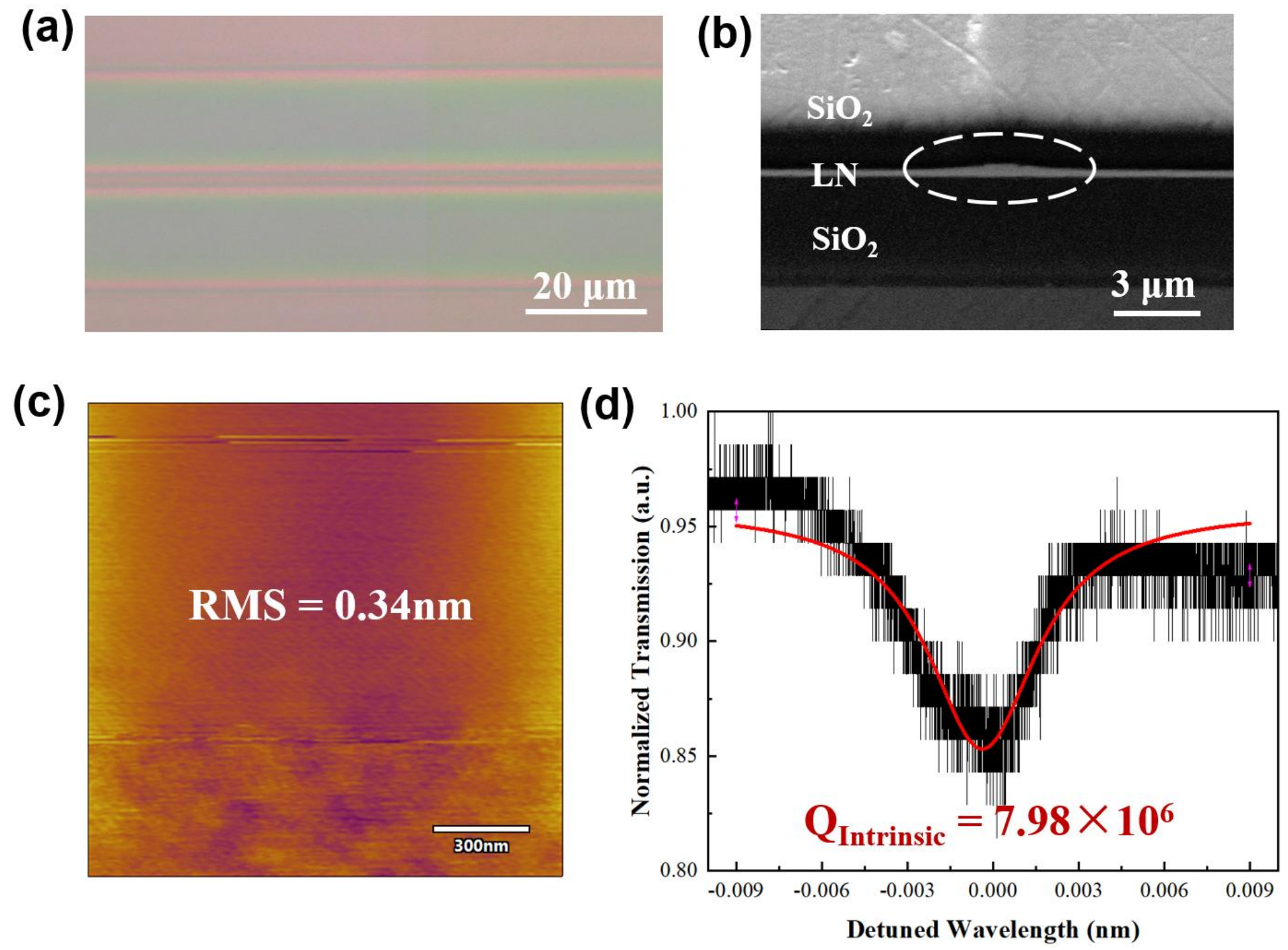

**Figure 2.** (a) Optical micrograph of the waveguide. (b) Cross-sectional SEM image. (c) AFM image, showing an RMS roughness of 0.34 nm. (d) Mode resonance of a racetrack microring resonator fabricated on the TFLN wafer by the same process, showing an intrinsic Q factor $Q_{int}$ of $7.98\times10^6$, corresponding to a propagation loss of only 0.042 dB/cm.

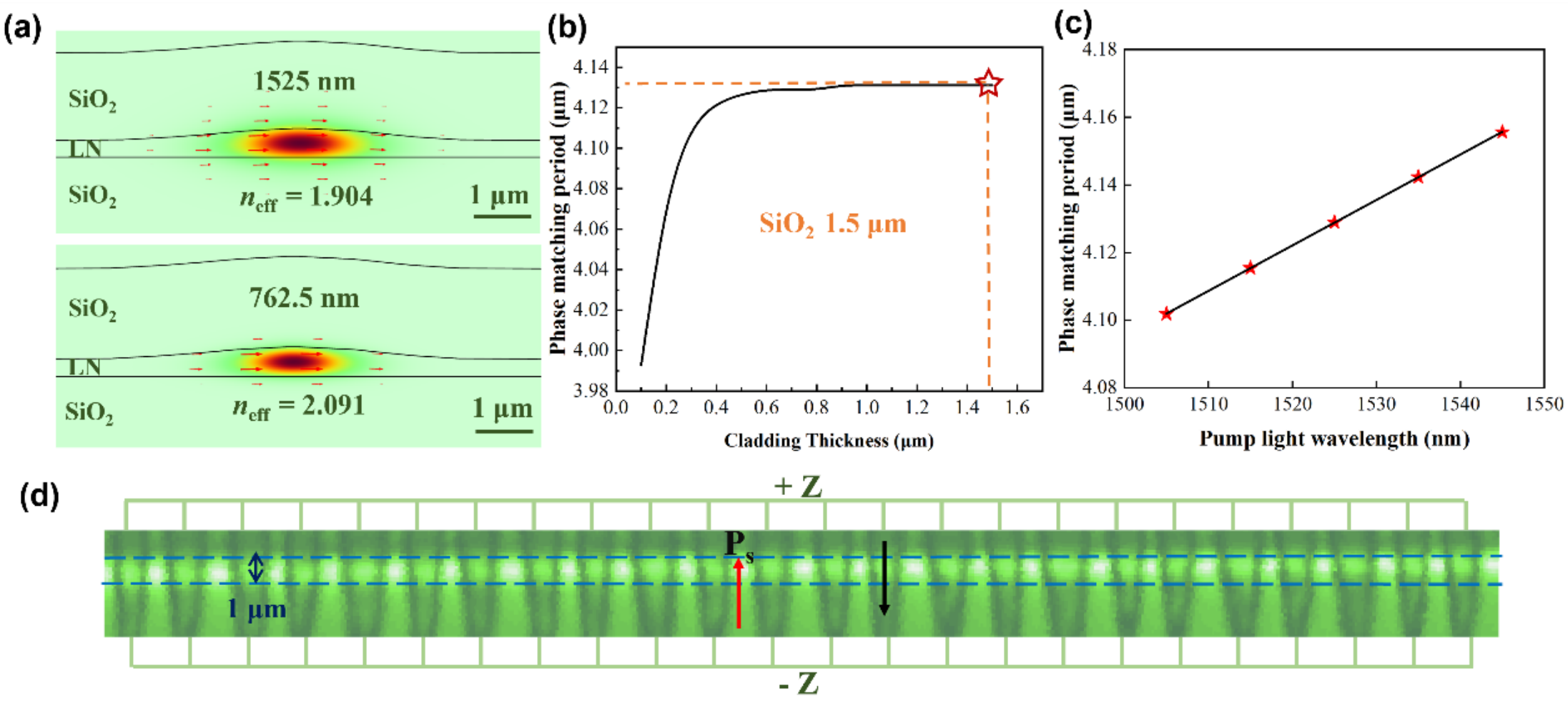

**Figure 3.** (a) Simulated $TE_{00}$-like fundamental and second-harmonic modes. (b) QPM period versus $SiO_2$ cladding thickness. (c) QPM period versus pump wavelength for a 1.5-μm-thick cladding. (d) SH microscopy image of periodic domains over approximately 100 μm.

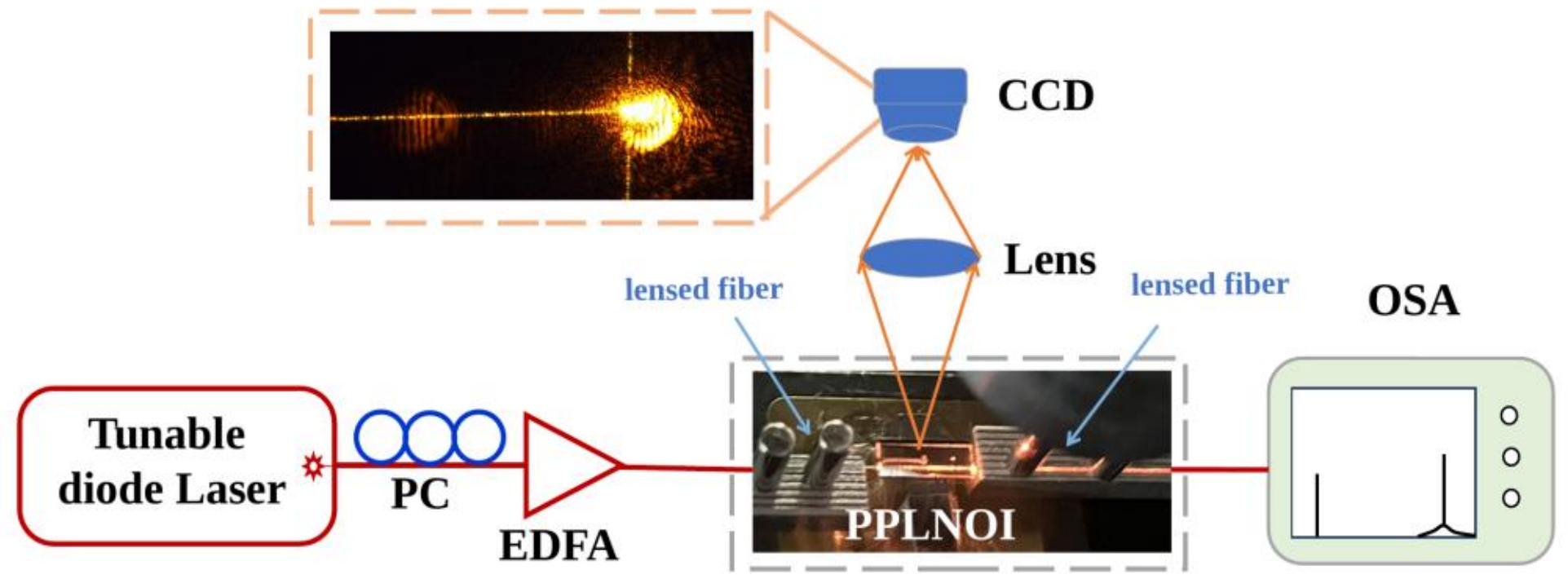

**Figure 4.** Experimental setup for SHG measurement. PC, polarization controller; EDFA, erbium-doped fiber amplifier; OSA, optical spectrum analyzer. Upper Inset: Optical microscope image of second harmonic emission from the output port the waveguide. Lower Inset: a photo of the waveguide under test, where PPLNOI denotes the periodically-poled-lithium-niobate-on-insulator chip.

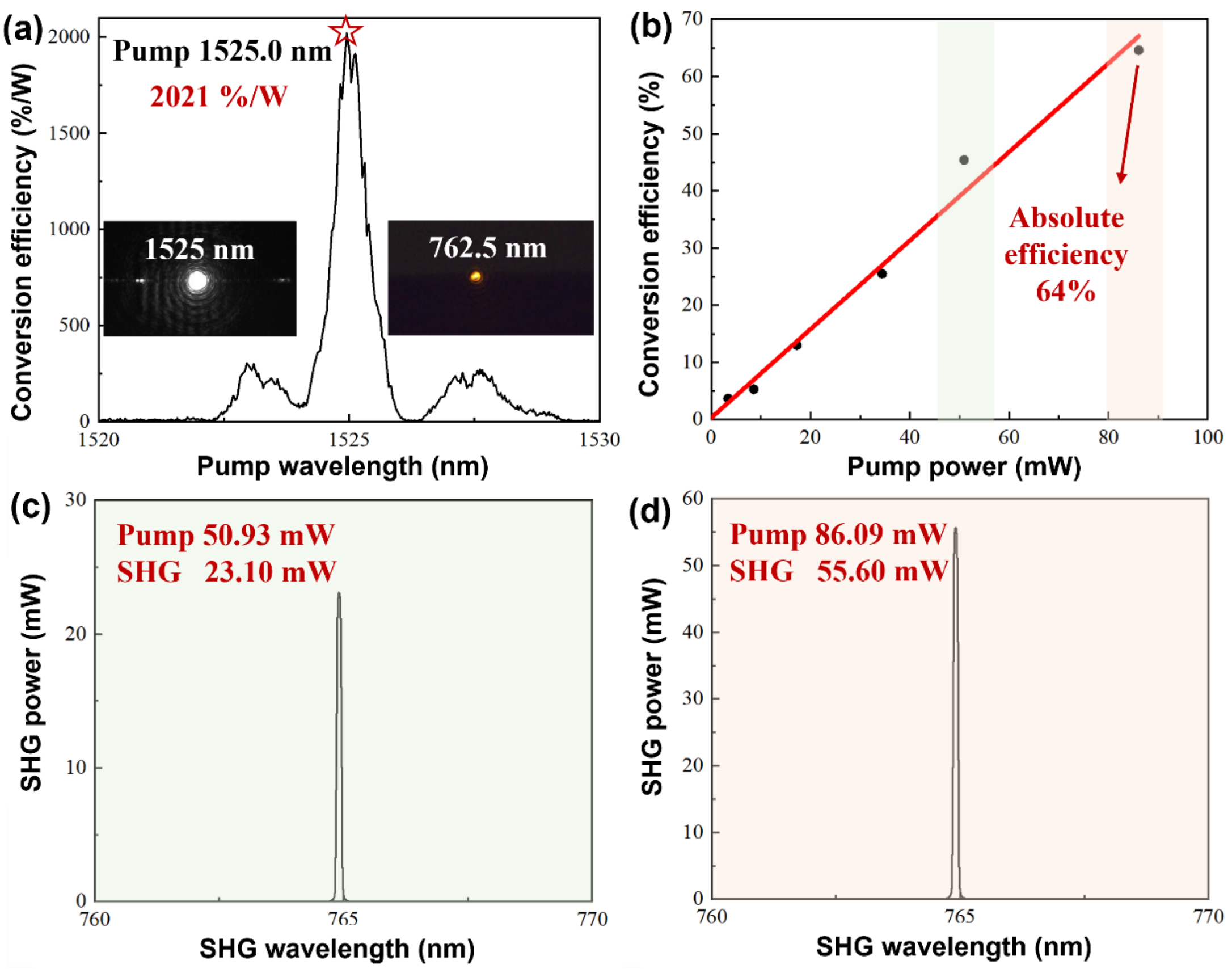

**Figure 5**. (a) Normalized conversion efficiency versus pump wavelength, showing 2021%/W near 1525 nm. (b) Absolute conversion efficiency versus on-chip pump power. SH spectra at approximately (c) 50-mW and (d) 86-mW pump powers.